\documentclass[a4paper,10pt]{article}

\usepackage{verbatim}
\usepackage{amsmath,amssymb}
\usepackage{enumerate}
\usepackage{color}

\newtheorem{defi}{Definition}%[section]
\newtheorem{lem}{Lemma}

\newtheorem{theo}{Theorem}

\newtheorem{rmq}{Remark}
\newtheorem{rmqs}[rmq]{Remarks}
\newenvironment{dem}{{\bf Proof }}{ { \hspace {\stretch{1} }$\Box$}}

\renewcommand{\d}{\mathrm{d}}

\renewcommand{\div}{\mathrm{div}}

\newcommand{\N}{{\mathbb N}}

\newcommand{\R}{{\mathbb R}}
\newcommand{\C}{{\mathbb C}}

\newcommand{\Aa}{{\mathcal A}}
\newcommand{\Bb}{{\mathcal B}}
\newcommand{\Cc}{{\mathcal C}}

\newcommand{\Ff}{{\mathcal F}}

\newcommand{\Ii}{{\mathcal I}}

\newcommand{\Ss}{{\mathcal S}}

\newcommand{\V}{{\mathcal V}}

\newcommand{\Y}{{\mathcal Y}}
\newcommand{\eps}{\varepsilon}
\renewcommand{\leq}{\leqslant}
\renewcommand{\geq}{\geqslant}

\newcommand{\h}{\eps}
\newcommand{\dt}{\frac{d}{dt}}
\newcommand{\kt}{{\kappa^t}}

\begin{document}

\title{\bf Semiclassical Simple Initial Value Representations}
\author{Vidian ROUSSE \\{\it \small Universit\'e Paris 12  -- UMR 8050}}

\date{ }

\maketitle

{\small{\bf Abstract}: In this article, a class of Fourier Integral Operators which converge to the unitary group of the Schr\"odinger equation in semiclassical limit $\eps\to 0$ is constructed. The convergence is in the uniform operator norm and allows for a error bound $C_N\eps^{N+1}$ for any integer $N$ and extends to Ehrenfest timescaleswith bound $C_N\eps^{N+1-\rho}$ where $\rho$ can be made arbitrary small. In the chemical literature those approximations are known as simple Initial Value Representations.
}

%%%%%%%%%%%%%%%%%%%%%%%%%%%%%%%%%%%%%%%%%%%%%%%%%%%%%%%%%%
\section{Introduction}
%%%%%%%%%%%%%%%%%%%%%%%%%%%%%%%%%%%%%%%%%%%%%%%%%%%%%%%%%%

We study approximate solutions of the semiclassical time-dependent Schr\"odinger equation
\begin{equation}\label{eq:TDSE}
i\eps\dt\psi^\eps(t)=-\frac{\eps^2}{2}{\rm\Delta}\psi^\eps(t)+V(x)\psi^\eps(t),\qquad\psi^\eps(0)=\psi^\eps_0\in L^2(\R^d,\C)
\end{equation}
in the semiclassical limit $\eps\to 0$. The operator $H^\eps:=-\frac{\eps^2}{2}{\rm\Delta}+V(x)$ on the right-hand side of \eqref{eq:TDSE} is the so-called Hamiltonian, a self-adjoint operator on $L^2(\R^d)$ (under suitable assumptions on the potential $V$).
It is well-known that the solution of~\eqref{eq:TDSE} can be written as
\begin{equation*}
\psi^\eps(t)=e^{-\frac{i}{\eps}H^\eps t}\psi^\eps_0,
\end{equation*}
where the group of unitary operators $e^{-\frac{i}{\eps}H^\eps t}$ is defined by the spectral theorem.

The semiclassical parameter $\eps$ may be thought of as the quantum of action $\hbar$, but there are also situations, where $\eps$ has a different meaning. One example is provided by Born-Oppenheimer molecular dynamics, where equation~\eqref{eq:TDSE} describes the semiclassical motion of the nuclei of a molecule in the case of well-separated electronic energy surfaces and $\eps$ is the square root of the ratio of the electronic mass and the average nuclear mass. In this case, the $\eps$ in front of the time-derivative in~\eqref{eq:TDSE} is due to a rescaling of time $\tilde{t}=t/\eps$. This particular choice, the so-called ``distinguished limit'' (see~\cite{[Cole]}), produces the most interesting results in the semiclassical limit $\eps\to 0$.

To formulate our main result, we introduce the following class of Fourier Integral Operators (FIOs):
\begin{equation}\label{eq:FIO}
\Ii^\eps(\kappa^t; u)\phi(x):=\frac{1}{(2\pi\eps)^d}\int_{\R^{2d}} e^{\frac{i}{\eps}\Phi^{\kappa^t}(x,y,\eta)}
u(x,y,\eta) \phi(y)\:d\eta\:dy,
\end{equation}
where
\begin{itemize}
\item $\kappa^t(q,p)=\left(X^{\kappa^t}(q,p),\Xi^{\kappa^t}(q,p)\right)$ is a $C^1$-family of \emph{canonical transformations} of the classical phase space $T^*\R^d=\R^d\times\R^d$,
\item $S^{\kappa^t}(q,p)$ is the associated \emph{classical action} (see Definition \ref{def:action} for a precise definition) which reduces to
\begin{equation*}
S^{\kappa^t}(q,p)=\int\limits_0^t\left[ \dt X^{\kappa^\tau}(q,p)\cdot\Xi^{\kappa^\tau}(q,p)-(h\circ\kappa^\tau)(q,p)\right]\;d\tau
\end{equation*}
when $\kappa^t$ is the Hamiltonian flow associated with a Hamiltonian function $h$,
\item the complex-valued {\em phase function} is given by
\begin{equation} \label{eq:phase}
\Phi^{\kappa^t}(x,y,\eta)=
S^{\kappa^t}(y,\eta)+\Xi^{\kappa^t}(y,\eta)\cdot\left(x-X^{\kappa^t}(y,\eta)\right)+\frac{i}{2}\left|x-X^{\kappa^t}(y,\eta)\right|^2
\end{equation}
\item and the {\em symbol} $u$ is a smooth complex-valued function which is bounded with all its derivatives.
\end{itemize}

%For this class of operators, the authors previously established an $L^2$-boundedness result, see \cite{[RousseSwart]}.
The central results of this paper reads

$ $

\textbf{Theorem A }{\it For fixed $t$, under the assumption that the canonical transformation $\kt$ is biLipschitzian with bounded derivatives, the operator $\Ii^\eps(\kappa^t;u)$ is continuous $\Ss(\R^d)\to\Ss(\R^d)$ and extends to a continuous operator $L^2(\R^d)\to L^2(\R^d)$ with
\begin{equation*}
\Vert\Ii^\eps(\kappa^t;u)\Vert_{L^2\to L^2}\leq C(\kappa)\sum_{|\alpha_k|\leq d}\Vert\partial^{\alpha_1}_x\partial^{\alpha_2}_y\partial^{\alpha_3}_\eta u\Vert_{L^\infty}.
\end{equation*}
}

\textbf{Theorem B }{\it 
Let $e^{-\frac{i}{\eps} H^\eps t}$ be the propagator defined by the time-dependent Schr\"o\-dinger equation~\eqref{eq:TDSE} on the time-interval $[-T,T]$ with subquadratic potential $V\in C^\infty(\R^d,\R)$, i.e.
$\sup_{x\in\R^{d}}|\partial^\alpha_{x}V(x)|<\infty$ for all $\alpha\in\N^{d} \textrm{ with }|\alpha|\geq 2$. Then
\begin{equation*}
\sup_{t\in [-T,T]}\left\|e^{-\frac{i}{\eps} H^\eps t}-\Ii^\eps\left(\kappa^t; u\right)\right\|_{L^2\to L^2}\leq C(T)\eps,
\end{equation*}
where $\kappa^t=(X^{\kappa^t},\Xi^{\kappa^t})$ and $u$ are uniquely given as
\begin{itemize}
\item the flow associated with the classical Hamiltonian $h(x,\xi)=\frac12|\xi|^2+V(x)$
\begin{equation*}
\left\{\begin{array}{rcl} \dt X^{\kappa^t}(q,p) & = & \Xi^{\kappa^t}(q,p) \\ \dt\Xi^{\kappa^t}(q,p) & = & -\nabla V\left(X^{\kappa^t}(q,p)\right) \end{array}\right.
\qquad
\left\{\begin{array}{rcl} X^{\kappa^0}(q,p) & = & q \\ \Xi^{\kappa^0}(q,p) & = & p \end{array}\right.
\end{equation*}
and
\item the solution of the Cauchy-problem
\begin{eqnarray*}
\dt u(t,y,\eta) & = & \frac12{\rm tr}\left[\Y_0\left(F^\kt(y,\eta)\right)^{-1}\frac{d}{dt}\Y_0\left(F^\kt(y,\eta)\right)\right]u(t,y,\eta) \\
u(0,y,\eta) & = & 1.
\end{eqnarray*}
\end{itemize}
The $\C^{d\times d}$-valued function
\begin{equation*}
\Y_0\left(F^{\kappa^t}(y,\eta)\right)=\Xi^{\kappa^t}_\eta(y,\eta)-i X^{\kappa^t}_\eta(y,\eta),
\end{equation*}
depends on elements of the transposed Jacobian
\begin{equation*}
F^{\kappa^t}(y,\eta)^\dagger=
\begin{pmatrix}
X^{\kappa^t}_y(y,\eta)&\Xi^{\kappa^t}_y(y,\eta)\\
X^{\kappa^t}_\eta(y,\eta)&\Xi^{\kappa^t}_\eta(y,\eta)\\
\end{pmatrix}
\end{equation*}
of $\kappa^t$ with respect to $(y,\eta)$.
}

The equation for $u$ is easily solved. Its solution
\begin{equation*}
u(t,y,\eta)=\left(\det\Y\left(F^{\kappa^t}(y,\eta)\right)\right)^\frac12,
\end{equation*}
(where the branch of the square root is chosen by continuity in time starting from $t=0$) mirrors the so-called Herman-Kluk prefactor as presented in \cite{[SwartRousse]}. The separation of the results in two steps is made on purpose to emphasize that the first $L^2$-boundedness result is a key argument into the proof of the second result. By the way, Theorem B is just a simplified version of our main result in Section \ref{s:evol}. More precisely, Theorem~\ref{theo:main_result} will essentially add two central aspects. First, for the Ehrenfest-timescale $T(\eps)=C_T\log(\eps^{-1})$ the result still holds with a slightly weaker bound. Second, the error estimate can be improved to $\eps^N$, where $N$ is arbitrary large by adding a correction of the form
$\sum_{n=1}^{N-1}\eps^n u_n$ to $u$. As $u$, the $u_n$ are solutions of explicitly solvable Cauchy-problems. Finally the opportunity to extend this result to general pseudodifferential operators will be discussed in Remark \ref{rk:toPDO}.

Whereas there is an abundant number of works on Fourier Integral Operators in the mathematical literature, only few of them discuss the relation between FIOs and the time-dependent Schr\"odinger-equation. The first works which apply FIOs with real-valued phase function to this problem are~\cite{[KitadaKumanoGo]} and~\cite{[Kitada]}. In this case one has to deal with the boundary value problem
$$
\textrm{Given }x,y\in\R^d\textrm{, find }p\textrm{ such that }X^{\kappa^t}(y,p)=x.
$$
To get uniqueness for its solution one has either to restrict to short times $t$ or to impose very strong restrictions on the potential. The same problems are met in~\cite{[Fujiwara1]}, where Fujiwara applies a related class of operators without integral in the oscillatory kernel to the Schr\"odinger equation to justify the time-slicing approach for Feynman's path integrals.

The avoidance of this problem is the major advantage of complex-valued phase functions. In the non-semiclassical setting, Tataru shows in \cite{[Tataru]} that the unitary group of time evolution is an FIO with complex-valued phase function (different from~\eqref{eq:phase} because his kernel consists of an integral over the whole phase-space space in contrast to the momentum integral we proposed). He also establishes that the use of constant symbols leads to a parametrix for the non-semiclassical Schr\"odinger equation. In the semiclassical setting, \cite{[BilyRobert]}, \cite{[RousseSwart]} and \cite{[SwartRousse]} treat FIOs of the same type (phase-space integral kernel) with closer connection to Gaussian wave packets known in the chemical literature as Initial Value Representations.

The class of operators defined in~\eqref{eq:FIO} is used in the works~\cite{[LaptevSigal]} and~\cite{[Butler]} for the construction of approximate solutions of the semiclassical time-dependent Schr\"o\-dinger equation. However, these works only allow compactly supported symbols, which enforces the truncation of the Hamiltonian at least in momentum. In particular, in~\cite{[LaptevSigal]} the comments of Theorem 2.1 mentioned that the dependence of a constant analog to our $C(T)$ on
\begin{itemize}
\item the size of the compact support,
\item the length of the time interval $T$ and
\item growth of a possible subprincipal symbol for $H^\eps$
\end{itemize}
was not controlled. The result presented here proposes a (partial) answer to those questions.

In contrast to the mathematical literature connecting time-dependent Schr\"o\-dinger equation and Fourier Integral Operators, there is an abundant number of papers in chemical journals on this topic. The reader interested in those aspects can consult \cite{[Kay]} (and references therein) which gives a good review of the different approximations used by chemists.

\subsection*{Organisation of the paper and notation}

The paper is organised in the following way. Section~\ref{s:FIOs} will introduce the class of FIOs with complex quadratic phase and constant spreading matrix $\Theta$. In Section~\ref{s:C0S}, we will prove their continuity on the Schwartz space $\Ss(\R^d)$ and explain how to push the dependence of the symbol on the variable $x$ to higher order in $\eps$. Section~\ref{s:L2bound} will adress the problem of the $L^2$-boundedness for those FIOs. The extension of all those results to non-constant spreading matrix $\Theta(y,\eta)$ will be hinted in Section~\ref{s:spread}. Finally, all these results will lead to our main result, which we will state in Theorem~\ref{theo:main_result} of Section~\ref{s:evol}.

We close this introduction by a short discussion of the notation. Throughout this paper, we will use standard multiindex notation. Vectors will always be considered as column vectors. The inner product of two vectors $a, b\in\R^d$ will be denoted as $a\cdot b=\sum_{j=1}^da_j b_j$ and extended to vectors $a,b\in\C^d$ by the same formula. The transpose of a matrix $A$ will be $A^\dagger$, whereas $A^*:=\bar{A}^\dagger$ denotes the adjoint and finally $e_j$ will stand for the $j$th canonical basis vector of $\R^d$ or $\C^d$.

For a differentiable mapping $F\in C^1(\R^d,\C^d)$, we will use both $(\partial_x F)(x)$ and $F_x(x)$ for the transpose of its Jacobian at $x$, i.e. $((\partial_x F)(x))_{jk}=(F_x(x))_{jk}=(\partial_{x_j} F_k)(x)$. This leads to the identity $\partial_x (F\cdot G)=G_x F+F_x G$ for $F,G\in C^1(\R^d,\C^d)$. The Hessian matrix of a mapping $F\in C^2(\R^d,\C)$ will be denoted by $\textrm{Hess}_{x}F(x)$.

For the sake of better readability of the formulae, we will be somewhat sloppy with respect to the distinction between functions and their values. As a crucial example, we will write $(x-X^\kappa(y,\eta))v$ for the function $(x,y,\eta)\mapsto(x-X^\kappa(y,\eta))v(x,y,\eta)$.

Finally, we remind the normalization for the Fourier transform we will be using
\begin{equation*}
(\Ff\varphi)(\xi)=\frac{1}{(2\pi)^{d/2}}\int_{\R^d}e^{-i\xi\cdot x}\varphi(x)\d x \qquad {\rm and} \qquad
(\Ff^{-1}\psi)(x)=\frac{1}{(2\pi)^{d/2}}\int_{\R^d}e^{i\xi\cdot x}\psi(\xi)\d\xi.
\end{equation*}

%%%%%%%%%%%%%%%%%%%%%%%%%%%%%%%%%%%%%%%%%%%%%%%%%%%%%%%%%%
\section{%Symbol Classes, 
%Canonical Transformations %, Action and Class $\Bb$
Semiclassical Fourier Integral Operators% with Complex Phase
\label{s:FIOs}}
%%%%%%%%%%%%%%%%%%%%%%%%%%%%%%%%%%%%%%%%%%%%%%%%%%%%%%%%%%

Our FIOs involve two fundamental objects: their symbol $u$ and a canonical transformation $\kappa$. We will first give more precision to the restrictions we put on those objects.

\begin{defi}[Symbol class]
Let ${\bf d}=(d_j)_{1\leq j\leq J}\in\N^J$, $u(z_1,\ldots,z_J)$ a smooth function of $\Cc^\infty(\R^{d_1}\times\cdots\times\R^{d_J};\C^N)$ and ${\bf m}=(m_j)_{1\leq j\leq J}\in\R^J$. We say that $u$ is a {\bf symbol of class $S[{\bf m};{\bf d}]$} if the following quantities are finite for any $k\geq0$
\begin{equation*}
M^m_k[u]:=\max_{\sum_{j=1}^J\alpha_j=k}\sup_{z_j\in\R^{d_j}}\left|\left(\prod_{j=1}^J\langle z_j\rangle^{-m_j}\partial^{\alpha_j}_{z_j}\right)u(z_1,\ldots,z_J)\right|
\end{equation*}
where $\langle z\rangle:=\sqrt{1+|z|^2}$.

We extend this definition to any $m_j\in\overline{\R}:=\{-\infty\}\cup\R\cup\{+\infty\}$ by setting, for instance with non-finite $m_1$,
\begin{equation*}
S[(+\infty,m_2,\ldots,m_J);{\bf d}]=\bigcup_{m_1\in\R}S[(m_1,\ldots,m_J);{\bf d}]
\end{equation*}
and
\begin{equation*}
S[(-\infty,m_2,\ldots,m_J);{\bf d}]=\bigcap_{m_1\in\R}S[(m_1,\ldots,m_J);{\bf d}]
\end{equation*}
and so on.
\end{defi}

To fix notations, we also recall the definition of a canonical transformation and the link with symplectic matrices.

\begin{defi}[Canonical transformation] \label{def:CanonicalTransform} $ $

\noindent Let $\kappa(q,p)$ be a smooth diffeomorphism from $\R^d\times\R^d$ into itself decomposed into position/momentum variables through $\kappa(q,p)=(X^\kappa(q,p),\Xi^\kappa(q,p))$. We represent its differential by the following Jacobian matrix
\begin{equation} \label{eq:Jacobian}
F^\kappa(q,p)=\left(\begin{array}{cc} X^\kappa_q(q,p)^\dagger & X^\kappa_p(q,p)^\dagger \\ \Xi^\kappa_q(q,p)^\dagger & \Xi^\kappa_p(q,p)^\dagger \end{array}\right).
\end{equation}
$\kappa$ is said to be a {\bf canonical transformation} if $F^\kappa(q,p)$ is symplectic for any $(q,p)$ in $\R^d\times\R^d$ {\it i.e.}
\begin{equation*}
[F^\kappa(q,p)]^\dagger JF^\kappa(q,p)=J \qquad {\rm where} \qquad J:=\left(\begin{array}{cc} 0 & I \\ -I & 0 \end{array}\right).
\end{equation*}
\end{defi}

As discussed in \cite{[RousseSwart]} and \cite{[SwartRousse]}, one can associate to a canonical transformation a real-valued function that plays the role of the Lagrangian action integral for Hamiltonian flows.

\begin{defi}[Action] \label{def:action}
Let $\kappa(q,p)=(X^\kappa(q,p),\Xi^\kappa(q,p))$ be a canonical transformation of $\R^d\times\R^d$. A real-valued function $S^\kappa$ is called an {\bf action associated with $\kappa$} if it fulfills
\begin{equation} \label{eq:action}
S^\kappa_q(q,p)=-p+X^\kappa_q(q,p)\Xi^\kappa(q,p) ,\quad S^\kappa_p(q,p)=X^\kappa_p(q,p)\Xi^\kappa(q,p).
\end{equation}
\end{defi}

\begin{rmq}
The function $S^\kappa$ always exists and is uniquely defined up to an additive constant.
\end{rmq}

We specialize here the notion of diffeomorphism of class $\Bb$ as presented by Fujiwara in~\cite{[Fujiwara1]}.

\begin{defi}
A canonical transformation $\kappa$ of $\R^d\times\R^d$ is said to be {\bf of class $\Bb$} if $F^\kappa\in S[0;2d]$ and for any $k\geq0$, we set $M^\kappa_k:=M^1_k[F^\kappa]$.
\end{defi}

\begin{rmq}
The subset of canonical transformations of class $\Bb$ is a subgroup (for composition) of bilipschitzian diffeomorphisms of $\R^d\times\R^d$ {\it i.e.} if $\kappa$ is a canonical transformation of class $\Bb$, then there exist two strictly positive constants $c_\kappa$ and $C_\kappa$ such that for any $(q_1,p_1)$ and $(q_2,p_2)$ in $\R^d\times\R^d$
\begin{equation} \label{biLipschitz}
c_\kappa\Vert(q_2,p_2)-(q_1,p_1)\Vert\leq\Vert\kappa(q_2,p_2)-\kappa(q_1,p_1)\Vert\leq C_\kappa\Vert(q_2,p_2)-(q_1,p_1)\Vert.
\end{equation}
\end{rmq}

$ $

From now on, all canonical transformations considered are assumed to be of class $\Bb$ and $\h$ will denote a small parameter such that $0<\h\leq1$.

%\subsection{Definitions and $\Ss$ Continuity}
%%%%%%%%%%%%%%%%%%%%%%%%%%%%%%%%%%%%%%%%%%%%%%%%%%%%%%%%%%

We turn now to the definition of the main object of this article: semiclassical FIOs with quadratic complex phase.

\begin{defi}[FIO]
Let $\Theta$ be a complex symmetric matrix with positive definite real part. For $u\in S[+\infty;3d]$ and $\varphi\in\Ss(\R^d;\C)$, we define the action on $\varphi$ of the {\bf semiclassical FIO associated with $\kappa$ with symbol $u$} as the oscillatory integral (see Appendix \ref{oscill})
\begin{equation*}
[\Ii^\h(\kappa;u;\Theta)\varphi](x):=\frac{1}{(2\pi\h)^d}\int_{\R^{2d}}\hspace{-0.3cm}e^{\frac{i}{\h}\Phi^\kappa(x,y,\eta;\Theta)}\left[u(x,y,\eta)\varphi(y)\right]\d\eta\d y
\end{equation*}
where $\Phi^\kappa$ is a complex-valued phase function given by
\begin{equation*}
\Phi^\kappa(x,y,\eta;\Theta)=S^\kappa(y,\eta)+\Xi^\kappa(y,\eta)\cdot(x-X^\kappa(y,\eta))+\frac{i}{2}(x-X^\kappa(y,\eta))\cdot\Theta(x-X^\kappa(y,\eta)).
\end{equation*}
%$L_y$ is the first order differential operator
%\begin{equation*}
%L_y=\frac{1}{1+|\nabla_y\Phi^\kappa(x,y,\eta;\Theta)|^2}\left[1-i\h\nabla_y\overline{\Phi^\kappa(x,y,\eta;\Theta)}\cdot\nabla_y\right]
%\end{equation*}
%and $L_y^\dagger$ stands for its symmetric given by
%\begin{equation*}
%\int_{\R^d}v(y)[L_y^\dagger u](y)\d y=\int_{\R^d}[L_yv](y)u(y)\d y.
%\end{equation*}
%If $m^\eta<-d$, its integral kernel is given by the absolutely convergent integral
%\begin{equation*}
%K^\h(\kappa;u;\Theta)(x,y):=\frac{1}{(2\pi\h)^d}\int_{\R^d}e^{\frac{i}{\h}\Phi^\kappa(x,y,\eta;\Theta)}u(x,y,\eta)\d\eta.
%\end{equation*}
\end{defi}

\begin{rmqs} \label{rk:rescaling}
$ $
\begin{enumerate}
%\item
%As already noticed in {\rm\cite{[AsadaFujiwara]}} or {\rm\cite{[Robert]}}, the following property can be alternatively used as a definition. If $\sigma\in\Ss(\R^d;\C)$ is such that $\sigma(0)=1$, we have
%\begin{equation} \label{defOscInt}
%[\Ii^\h(\kappa;u;\Theta)\varphi](x)=\lim_{\lambda\to+\infty}[\Ii^\h(\kappa;u^\lambda_\sigma;\Theta)\varphi](x) 
%\end{equation}
%where $u^\lambda_\sigma(x,y,\eta):=\sigma(\eta/\lambda)u(x,y,\eta)\in S[(+\infty,-\infty);(2d,d)]$.
\item
For any $u\in S[+\infty;3d]$, it results from the usual machinery of oscillatory integrals that the operator $\Ii^\h(\kappa;u;\Theta)$ is continuous from $\Ss(\R^d)$ into its dual $\Ss'(\R^d)$.
\item
The normalization is adjusted so that we have $\Ii^\h({\rm Id};1;\Theta)={\rm Id}$.
\item
With the rescalings
\begin{itemize}
\item
$T^\h_d\varphi(y):=\h^{d/4}\varphi(\sqrt{\h}y)$ (which is unitary on $L^2(\R^d)$),
\item
$\kappa^{(\h)}(y,\eta):=\kappa(\sqrt{\h}y,\sqrt{\h}\eta)/\sqrt{\h}$ (which preserves the symplectic structure),
\item
$u^{(\h)}(x,y,\eta):=u(\sqrt{\h}x,\sqrt{\h}y,\sqrt{\h}\eta)$,
\end{itemize}
we have $S^{\kappa^{(\h)}}(y,\eta)=S^\kappa(\sqrt{\h}y,\sqrt{\h}\eta)/\h$,
\begin{equation*}
\Phi^{\kappa^{(\h)}}(x,y,\eta;\Theta)=\Phi^\kappa(\sqrt{\h}x,\sqrt{\h}y,\sqrt{\h}\eta;\Theta)/\h
\end{equation*}
and
\begin{equation} \label{rescaling}
I^\h(\kappa;u;\Theta)=(T^\h_d)^*\Ii^1(\kappa^{(\h)};u^{(\h)};\Theta)T^\h_d.
\end{equation}
\item One might connect those FIOs to the one defined in {\rm\cite{[RousseSwart]}} and {\rm\cite{[SwartRousse]}} by taking the limit $\Theta^y\to+\infty$ in the latest with an appropriate renormalization. However, it turns out that this limit process does not allow to transfer their more interesting properties (essentially $L^2$-boundedness) to $\Ii^\h(\kappa;u;\Theta)$.
\end{enumerate}
\end{rmqs}

%Unfortunately, this definition is rather unsymmetric between departure and arrival variables and we will need a more symmetric object already studied in \cite{[RousseSwart]} and \cite{[SwartRousse]}.
%
%\begin{defi}[FIO $2$]
%Let $\Theta^x$ and $\Theta^y$ be two complex symmetric matrix with positive definite real part. For $u\in S[+\infty;4d]$ and $\varphi\in\Ss(\R^d;\C)$, we define the action of the {\bf semiclassical FIO associated with $\kappa$ with symbol $u$ on $\varphi$} as the oscillatory integral
%\begin{equation*}
%[\underline{\Ii}^\h(\kappa;u;\Theta^x,\Theta^y)\varphi](x):=\frac{1}{(2\pi\h)^{3d/2}}\int_{\R^{3d}}\hspace{-0.3cm}e^{\frac{i}{\h}\underline{\Phi}^\kappa(x,y,q,p;\Theta^x,\Theta^y)}\left[u(x,y,q,p)\varphi(y)\right]\d q\d p\d y
%\end{equation*}
%where $\underline{\Phi}^\kappa$ is a complex-valued phase function given by
%\begin{eqnarray*}
%\underline{\Phi}^\kappa(x,y,q,p;\Theta^x,\Theta^y) & = & S^\kappa(q,p)+\Xi^\kappa(q,p)\cdot(x-X^\kappa(q,p))-p\cdot(y-q) \\
% & & +\frac{i}{2}(x-X^\kappa(q,p))\cdot\Theta^x(x-X^\kappa(q,p))+\frac{i}{2}(y-q)\cdot\Theta^y(y-q).
%\end{eqnarray*}
%\end{defi}

%%%%%%%%%%%%%%%%%%%%%%%%%%%%%%%%%%%%%%%%%%%%%%%%%%%%%%%%%%
\section{$\Ss$ Continuity and Integration by Parts \label{s:C0S}}
%%%%%%%%%%%%%%%%%%%%%%%%%%%%%%%%%%%%%%%%%%%%%%%%%%%%%%%%%%

%We first recall the result obtained for operators $\underline{\Ii}^\h$ in \cite{[RousseSwart]} and \cite{[SwartRousse]}.
%
%\begin{theo} \label{C0Sold}
%If $u\in S[+\infty;4d]$, then $\underline{\Ii}^\h(\kappa;u;\Theta^x,\Theta^y)$ sends $\Ss(\R^d;\C)$ into itself and is continuous. Moreover, for finite $m^x$, $m^y$, $m^q$ and $m^p$, the map
%\begin{equation*}
%u\in S[(m^x,m^y,m^q,m^p);(d,d,d,d)]\mapsto(\underline{\Ii}^\h(\kappa;u;\Theta^x,\Theta^y):\Ss(\R^d;\C)\to\Ss(\R^d;\C))
%\end{equation*}
%is continuous.
%\end{theo}
%
%Analogously to that situation, one easily shows the following result.

The continuity obtained for the FIOs of \cite{[RousseSwart]} and \cite{[SwartRousse]} on the Schwartz space $\Ss(\R^d)$ can easily be obtained for $\Ii^\h(\kappa;u;\Theta)$ by the same type of argument.

\begin{theo} \label{C0S}
If $u\in S[+\infty;3d]$, then $\Ii^\h(\kappa;u;\Theta)$ sends $\Ss(\R^d)$ into itself and is continuous. Moreover, for finite $m^x$, $m^y$ and $m^\eta$, the map
\begin{equation*}
u\in S[(m^x,m^y,m^\eta);(d,d,d)]\mapsto(\Ii^\h(\kappa;u;\Theta):\Ss(\R^d)\to\Ss(\R^d))
\end{equation*}
is continuous.
\end{theo}

The proof follows from various integration by parts. We do not give a detailed proof here but remind the main tools.

We compute the derivatives of $\Phi^\kappa$ with respect to the variables $x$, $y$ and $\eta$
\begin{equation} \label{derivPhi}
\left(\begin{array}{c} \Phi^\kappa_x \\ \Phi^\kappa_y \\ \Phi^\kappa_\eta \end{array}\right)=
\left(\begin{array}{ccc} I & 0 & i\Theta \\ 0 & -I & \left[\Xi^\kappa_y-iX^\kappa_y\Theta\right](y,\eta) \\ 0 & 0 & \left[\Xi^\kappa_\eta-iX^\kappa_\eta\Theta\right](y,\eta) \end{array}\right)
\left(\begin{array}{c} \Xi^\kappa(y,\eta) \\ \eta \\ x-X^\kappa(y,\eta) \end{array}\right).
\end{equation}

In order to integrate by parts, we need the invertibility of this $3d\times3d$ matrix so we begin by establishing invertibility properties of
\begin{equation} \label{eq:Y}
\Y(F;\Theta):=\left(\begin{array}{cc} 0 & I \end{array}\right)F^\dagger\left(\begin{array}{c} -i\Theta \\ I \end{array}\right)=D^\dagger-iB^\dagger\Theta
\end{equation}
for a matrix $F$ with block decomposition $\left(\begin{array}{cc} A & B \\ C & D \end{array}\right)$.

\begin{lem} \label{Winv}
If $F$ is symplectic, $\Y(F;\Theta)$ is invertible and if $\kappa$ is of class $\Bb$, $\Y(F^\kappa(\cdot,\cdot);\Theta)^{-1}$ is in the class $S[0;2d]$.
\end{lem}

\begin{dem}
A straightforward computation shows that
\begin{equation*}
\Y(F;\Theta)\left(\Re\Theta\right)^{-1}\Y(F;\Theta)^*=\V(F;\Theta)^*\V(F;\Theta)
\end{equation*}
where
\begin{equation} \label{Wdefpos}
\V(F;\Theta)=\left(\begin{array}{cc} (\Re\Theta)^{-1/2}\Im\Theta & (\Re\Theta)^{-1/2} \\ (\Re\Theta)^{1/2} & 0 \end{array}\right)
F\left(\begin{array}{c} 0 \\ I \end{array}\right).
\end{equation}
Hence the invertibility of $\Y(F;\Theta)$ as the kernel of $\V(F;\Theta)$ is reduced to $\{0\}$.

If $\kappa$ is of class $\Bb$, then $\Y(F^\kappa;\Theta)$ is clearly in $S[0;2d]$ so the formula of the inverse with minors shows that it is enough to prove a bound from below for the determinant. $\Y(F;\Theta)$ is linear continuous in $F$ and, as $\kappa$ is of class $\Bb$, $F^\kappa$ has relatively compact image, hence $|\det\Y(F;\Theta)|$ reaching its minimum on this compact, minimum which can not be zero from above. 
\end{dem}

$ $

Thus, integrating by parts with respect to $\eta$ allows to convert any polynomial growth in $x$ to the variables $(y,\eta)$ through the decomposition $x=X^\kappa(y,\eta)+(x-X^\kappa(y,\eta))$. More precisely, the removal of $x-X^\kappa(y,\eta)$ induces a gain in $\h$ and is stated as follows.

\begin{lem} \label{push}
Let $u\in S[+\infty,3d]$ and $V(x,y,\eta)$ a $\C^d$-valued function in $S[+\infty,3d]$, then
\begin{equation} \label{eq:IPP}
\Ii^\h\left(\kappa;V(x,y,\eta)\cdot(x-X^\kappa(y,\eta))u;\Theta\right)=i\h \Ii^\h\left(\kappa;L(\kappa;\Theta;V)u;\Theta\right)
\end{equation}
where
\begin{equation} \label{IPP}
[L(\kappa;\Theta;V)u](x,y,\eta):=\div_\eta\left[u(x,y,\eta){\Y(F^\kappa(y,\eta);\Theta)^\dagger}^{-1}V(x,y,\eta)\right].
\end{equation}
%and $L(\kappa;\Theta;V):S[(m^x,m^y,m^\eta);(d,d,d)]\to S[(m^x,m^y,m^\eta);(d,d,d)]$ is continuous
\end{lem}

It remains to remove the possible polynomial growth in $\eta$ which follows from the relation
\begin{equation*}
\left(\nabla_y+\Aa(F^\kappa(y,\eta);\Theta)\nabla_\eta\right)\Phi^\kappa=-\eta
\end{equation*}
where $\Aa(F^\kappa(y,\eta);\Theta)$ stands for the $S[0;2d]$ matrix given by
\begin{equation} \label{Aa}
\Aa(F;\Theta)=\left[C^\dagger-iA^\dagger\Theta\right]\Y(F;\Theta)^{-1}=\left[C^\dagger-iA^\dagger\Theta\right]\left[D^\dagger-iB^\dagger\Theta\right]^{-1}.
\end{equation}

\begin{lem} \label{push2}
Let $u\in S[+\infty,3d]$ and $V(x,y,\eta)$ a $\C^d$-valued function in $S[+\infty,3d]$, then
\begin{equation*} \label{eq:IPP2}
\Ii^\h\left(\kappa;V(x,y,\eta)\cdot\eta u;\Theta\right)=-i\h \Ii^\h\left(\kappa;L'(\kappa;\Theta;V)u;\Theta\right)
\end{equation*}
where
\begin{equation*} \label{IPP2}
[L'(\kappa;\Theta;V)u](x,y,\eta):=\div_y\left[u(x,y,\eta)V(x,y,\eta)\right]+\div_\eta\left[u(x,y,\eta)\Aa(F^\kappa(y,\eta);\Theta)^\dagger V(x,y,\eta)\right].
\end{equation*}
%and $L(\kappa;\Theta;V):S[(m^x,m^y,m^\eta);(d,d,d)]\to S[(m^x,m^y,m^\eta);(d,d,d)]$ is continuous
\end{lem}

%%%%%%%%%%%%%%%%%%%%%%%%%%%%%%%%%%%%%%%%%%%%%%%%%%%%%%%%%%
\section{$L^2$ Continuity \label{s:L2bound}}
%%%%%%%%%%%%%%%%%%%%%%%%%%%%%%%%%%%%%%%%%%%%%%%%%%%%%%%%%%

In this section, we will prove an $L^2$-boundedness result for our FIOs analogous to the Calder\'on-Vaillancourt Theorem (see \cite{[CalderonVaillancourt]}) for pseudodifferential operators.% We assume that $0<\lambda I\leq\Re\Theta\leq\gamma I$ as quadratic forms (one can take $\gamma=\Vert\Re\Theta\Vert$ and $\lambda=\Vert(\Re\Theta)^{-1}\Vert^{-1}$).

\begin{theo} \label{theo:L2bound}
Let $u\in S[0;3d]$ be a symbol and $\kappa$ a canonical transformation of class $\Bb$, then $\Ii^\h(\kappa;u;\Theta)$ can be extended in a unique way to a linear bounded operator $L^2(\R^d)\to L^2(\R^d)$ and there exists $C(\kappa;\Theta)>0$ such that
\begin{equation} \label{eq:full}
\left\Vert\Ii^\h(\kappa;u;\Theta)\right\Vert_{L^2\to L^2}\leq C(\kappa;\Theta)\sum_{\substack{\alpha_j=0,1 \\ \beta_j=0,1 \\ |\gamma|\leq d}}%\h^{\frac{|\alpha|+|\beta|+|\gamma|}{2}}
\Vert\partial^\alpha_x\partial^\beta_y\partial^\gamma_\eta u\Vert_\infty.
\end{equation}
%where
%\begin{equation*}
%\Vert w\Vert_{W^{4d+1,\infty}_{(x,y)}L^\infty_{\eta}}:=\sum_{|\alpha|\leq4d+1}\Vert\partial^\alpha_{(x,y)}w\Vert_{L^\infty}.
%\end{equation*}
\end{theo}

%\begin{rmq}
%By duality, we have $\Ii^\h(\kappa;u;\Theta):\Ss'(\R^d;\C)\to\Ss'(\R^d;\C)$, thus, for $\varphi\in L^2(\R^d;\C)$, $\Ii^\h(\kappa;u;\Theta)\varphi$
%exists as a distribution in $\Ss'(\R^d;\C)$.
%
%Moreover,
%$$\left\langle\lambda^{d/2}\underline{\Ii}^\h(\kappa;u;\Theta,\lambda{\rm Id})\varphi|\psi\right\rangle
%\underset{\lambda\to+\infty}{\longrightarrow}\left\langle\Ii^\h(\kappa;u;\Theta)\varphi|\psi\right\rangle$$
%for $\varphi\in L^2(\R^d;\C)$ and $\psi\in\Ss(\R^d;\C)$ by dominated convergence.
%%$$\left|\left\langle\Ii^\h(\kappa;u;\Theta)\varphi|\psi\right\rangle\right|\leq
%%\left(\sup_{\lambda\geq1}\left\Vert\lambda^{d/2}\underline{\Ii}^\h(\kappa;u;\Theta,\lambda{\rm Id})\right\Vert\right).\Vert\varphi\Vert_{L^2}.\Vert\psi\Vert_{L^2}$$
%%so that $\Ii^\h(\kappa;u;\Theta)\varphi\in L^2(\R^d;\C)$ and $\Ii^\h(\kappa;u;\Theta)$ is continuous on $L^2(\R^d;\C)$ with
%%$$\Vert\Ii^\h(\kappa;u;\Theta)\Vert\leq\sup_{\lambda\geq1}\left\Vert\lambda^{d/2}\underline{\Ii}^\h(\kappa;u;\Theta,\lambda{\rm Id})\right\Vert.$$
%\end{rmq}

In the situation of a compactly supported symbol independent of $x$ already mentioned in \cite{[LaptevSigal]} and \cite{[Butler]}, it is rather straightforward to adapt the proof of Proposition~$5$ in \cite{[RousseSwart]} to get the following result.

\begin{theo}
Let $u\in\Cc^\infty_c(\R^{2d};\C)$ be a compactly supported symbol in $(y,\eta)$ and $\kappa$ a canonical transformation, then $\Ii^\h(\kappa;u;\Theta)$ can be extended in a unique way to a linear bounded operator $L^2(\R^d)\to L^2(\R^d)$ and there exists $C(\kappa;\Theta)>0$ such that
\begin{equation} \label{eq:full2}
\left\Vert\Ii^\h(\kappa;u;\Theta)\right\Vert_{L^2\to L^2}\leq C(\kappa;\Theta)\sum_{|\alpha|\leq d+1}\Vert\partial^\alpha_\eta u\Vert_\infty.
\end{equation}
\end{theo}

\begin{dem}{\bf of Theorem~\ref{theo:L2bound}}
We first note that the scalings of the third point of Remark \ref{rk:rescaling} indicate that it is enough to consider the case $\h=1$.
We will then use the strategy of Hwang in \cite{[Hwang]}.
We have
\begin{eqnarray*}
\left\langle\psi\mid\Ii^1(\kappa;u;\Theta)\varphi\right\rangle
 & = & \frac{1}{(2\pi)^d}\int_{\R^{3d}}e^{i\Phi^\kappa(x,y,\eta;\Theta)}u(x,y,\eta)\overline{\psi(x)}\varphi(y)\d y\d x\d\eta \\
 & = & \frac{1}{(2\pi)^{3d/2}}\int_{\R^{4d}}e^{i[\Phi^\kappa(x,y,\eta;\Theta)-\xi.y]}u(x,y,\eta)\overline{\psi(x)}(\Ff^{-1}\varphi)(\xi)\d\xi\d y\d x\d\eta.
\end{eqnarray*}
A straightforward computation shows that
\begin{equation*}
\left(\nabla_y+\Aa(F^\kappa(y,\eta);\Theta)\nabla_\eta\right)[\Phi^\kappa(x,y,\eta;\Theta)-\xi.y]=-(\xi+\eta)
\end{equation*}
where $\Aa(F;\Theta)$ is defined in \eqref{Aa}. Thus, one can integrate by parts $d$ times to write the integrand in $\langle\psi\mid\Ii^1(\kappa;u;\Theta)\varphi\rangle$ as
\begin{eqnarray*}
\lefteqn{e^{i[\Phi^\kappa(x,y,\eta;\Theta)-\xi.y]}\left[\prod_{j=1}^d\frac{1+i\partial_{y_j}+i\sum_{k=1}^d\partial_{\eta_k}\circ\Aa_{jk}(F^\kappa(y,\eta);\Theta)}{1+i(\xi_j+\eta_j)}\right]u(x,y,\eta)\overline{\psi(x)}(\Ff^{-1}\varphi)(\xi)} \\
 & = & \hspace{-.5cm}\sum_{\substack{\beta_j=0,1 \\ |\gamma|\leq d \\ |\delta|\leq d}}e^{i[\Phi^\kappa(x,y,\eta;\Theta)-\xi.y]}h_{\beta\gamma\delta}(y,\eta;\Theta)\left[\partial^\beta_y\partial^\gamma_\eta u(x,y,\eta)\right]\overline{\psi(x)}\prod_{j=1}^d\frac{1}{(1+i(\xi_j+\eta_j))^{1+\delta_j}}(\Ff^{-1}\varphi)(\xi)
\end{eqnarray*}
with $h_{\beta\gamma\delta}\in S[0,2d]$. Hence, by Fubini theorem
\begin{equation*}
\left\langle\psi\mid\Ii^1(\kappa;u;\Theta)\varphi\right\rangle=
\sum_{\substack{\beta_j=0,1 \\ |\gamma|\leq d \\ |\delta|\leq d}}\int_{\R^{2d}}e^{i[S^\kappa-\Xi^\kappa.X^\kappa](y,\eta)}h_{\beta\gamma\delta}(y,\eta;\Theta)\overline{\Psi_{\beta\gamma}(\kappa(y,\eta);\kappa;\Theta)}\Phi_\delta(y,\eta)\d y\d\eta
\end{equation*}
where we have set
\begin{eqnarray*}
\Psi_{\beta\gamma}(X,\Xi;\kappa;\Theta) & = & \int_{\R^d}e^{-i\Xi\cdot x}\psi(x)\overline{e^{-\frac{1}{2}(x-X)\cdot\Theta(x-X)}(\partial^\beta_y\partial^\gamma_\eta u)(x,\kappa^{-1}(X,\Xi))}\d x \\
\Phi_\delta(y,\eta) & = & \frac{1}{(2\pi)^{3d/2}}\int_{\R^d}e^{-i\xi\cdot y}(\Ff^{-1}\varphi)(\xi)\overline{\prod_{j=1}^d\frac{1}{(1-i(\xi_j+\eta_j))^{1+\delta_j}}}\d\xi.
\end{eqnarray*}
Now, we will interprete those functions as the short-time Fourier transform (or cross-Wigner distribution) of $L^2$ functions and use the results and notations of Appendix \ref{Wigner}. Precisely, we have (see Definition \ref{def:stft} and formula \eqref{stft2})
\begin{equation*}
\Psi_{\beta\gamma}(X,\Xi;\kappa;\Theta)=\tilde{V}_{g_{\beta\gamma}}[\psi](-X,\Xi) \qquad {\rm and} \qquad \Phi_\delta(y,\eta)=\frac{1}{(2\pi)^{3d/2}}V_{p_\delta}[\Ff^{-1}\varphi](y,-\eta)
\end{equation*}
where
\begin{equation} \label{p}
g_{\beta\gamma}(x,X,\Xi;\Theta)=e^{-\frac{1}{2}x\cdot\Theta x}(\partial^\beta_y\partial^\gamma_\eta u)(x+X,\kappa^{-1}(X,\Xi)) \quad {\rm and} \quad
p_\delta(\xi)=\prod_{j=1}^d\frac{1}{(1-i\xi_j)^{1+\delta_j}}.
\end{equation}
Thus,
\begin{eqnarray*}
\left\langle\psi\mid\Ii^1(\kappa;u;\Theta)\varphi\right\rangle
 & \leq & \sum_{\substack{\beta_j=0,1 \\ |\gamma|\leq d;\ |\delta|\leq d}}\hspace{-.3cm}\Vert h_{\beta\gamma\delta}\Vert_{L^\infty}.\Vert\Psi_{\beta\gamma}\Vert_{L^2}.\Vert\Phi_\delta\Vert_{L^2} \\
 & \leq & \sum_{\substack{\alpha_j=0,1;\ \beta_j=0,1 \\ |\gamma|\leq d;\ |\delta|\leq d}}\hspace{-.4cm}\Vert h_{\beta\gamma\delta}\Vert_{L^\infty}C(\Theta)\Vert\partial^\alpha_x\partial^\beta_y\partial^\gamma_\eta u\Vert_{L^\infty}\Vert\psi\Vert_{L^2}\frac{1}{(2\pi)^{3d/2}}\Vert p_\delta\Vert_{L^2}\Vert\varphi\Vert_{L^2} \\
 & \leq & \frac{\pi^{d/2}C(\Theta)}{(2\pi)^{3d/2}}\Vert\psi\Vert_{L^2}\Vert\varphi\Vert_{L^2}\sum_{\substack{\alpha_j=0,1;\ \beta_j=0,1 \\ |\gamma|\leq d;\ |\delta|\leq d}}\hspace{-.3cm}\Vert h_{\beta\gamma\delta}\Vert_{L^\infty}\Vert\partial^\alpha_x\partial^\beta_y\partial^\gamma_\eta u\Vert_{L^\infty}
\end{eqnarray*}
which closes the proof.
\end{dem}

%%%%%%%%%%%%%%%%%%%%%%%%%%%%%%%%%%%%%%%%%%%%%%%%%%%%%%%%
\section{From Frozen to Thawed Gaussians \label{s:spread}}
%%%%%%%%%%%%%%%%%%%%%%%%%%%%%%%%%%%%%%%%%%%%%%%%%%%%%%%%

As presented before, the complex matrix $\Theta$ does not depend on $(x,y,\eta)$. For applications, it might be useful to allow such a dependence. Obviously, as $\Theta$ should control the spreading of the Gaussian in $x$ around the phase-space point $\kappa(y,\eta)$, it is rather unadequate to allow for $x$-dependence (otherwise it is not a Gaussian anymore). Thus, we will treat the situation where $\Theta(y,\eta)$ only depends on $y$ and $\eta$. The following definition gives a precise idea of which dependence is allowed.

\begin{defi}[Admissible spreading]
Let $\Theta$ be a function on $\R^d\times\R^d$ with values in the set of complex symmetric matrices of positive definite real part. We will say that $\Theta$ is an {\bf admissible spreading matrix} ($\Theta\in\Cc$) if $\Theta\in S[0;2d]$ and there exists a constant positive definite real symmetric matrix $\Theta_0$ such that
\begin{equation*}
\forall(y,\eta)\in\R^d\times\R^d, \quad \forall V\in\R^d, \quad V\cdot\Re\Theta(y,\eta)V\geq V\cdot\Theta_0V. 
\end{equation*}
\end{defi}

The definition of FIOs for those general admissible spreading matrices is mostly the same.

\begin{defi}[FIO2]
Let $\Theta$ be an admissible spreading matrix in $\Cc$. For $u\in S[+\infty;3d]$ and $\varphi\in\Ss(\R^d;\C)$, we define the action on $\varphi$ of the {\bf semiclassical FIO associated with $\kappa$ with symbol $u$} as the oscillatory integral
\begin{equation*}
[\Ii^\h(\kappa;u;\Theta)\varphi](x):=\frac{1}{(2\pi\h)^d}\int_{\R^{2d}}\hspace{-0.3cm}e^{\frac{i}{\h}\Phi^\kappa(x,y,\eta;\Theta(y,\eta))}\left[u(x,y,\eta)\varphi(y)\right]\d\eta\d y.
\end{equation*}
\end{defi}

Let us now justify that Theorems \ref{C0S} and \ref{theo:L2bound} still holds for those FIOs. The key idea, as presented in \cite{[Swart]}, is to observe that the non-constant spreading part could be driven into the symbol without any damage. More precisely, we will split the Gaussian into
\begin{equation*}
\exp\left(-\frac{1}{2\eps}x\cdot\Theta(y,\eta)x\right)=\exp\left(-\frac{1}{2\eps}x\cdot\frac{\Theta_0}{2}x\right)\exp\left(-\frac{1}{2\eps}x\cdot\left[\Theta(y,\eta)-\frac{\Theta_0}{2}\right]x\right).
\end{equation*}
Thus, we have
\begin{equation*}
\Ii^\h(\kappa;u;\Theta)=\Ii^\h(\kappa;v^\eps[\Theta];\Theta_0)
\end{equation*}
with the new symbol
\begin{equation*}
v^\eps[\Theta](x,y,\eta)=u(x,y,\eta)\exp\left(-\frac{1}{2\eps}(x-X^\kappa(y,\eta))\cdot\left[\Theta(y,\eta)-\frac{\Theta_0}{2}\right](x-X^\kappa(y,\eta))\right)
\end{equation*}
which is $S[+\infty,3d]$ if $u$ does (the polynomial growth appearing when differentiating are compensated by the Gaussian decay with spreading still $\Theta_0/2$). As for the extension of Theorem~\ref{theo:L2bound}, it is enough to notice that the scalings of the third point of Remark \ref{rk:rescaling} implies a stronger result where each partial derivative of \eqref{eq:full} induces a gain of $\sqrt{\h}$ which compensates the $\sqrt{\h}^{-1}$ that appears when differentiating the symbol $v^\eps[\Theta]$.

%%%%%%%%%%%%%%%%%%%%%%%%%%%%%%%%%%%%%%%%%%%%%%%%%%%%%%%%%
%\section{Composition with Pseudodifferential Operators}
%%%%%%%%%%%%%%%%%%%%%%%%%%%%%%%%%%%%%%%%%%%%%%%%%%%%%%%%%
%
%\fbox{complete}
%${\rm Op}^\h h\circ\Ii^\h(\kappa;u;\Theta)=\Ii^\h(\kappa;(h\circ\kappa)u;\Theta)+\eps\cdots$
%
%\begin{equation*}
%[{\rm Op}^\h\varphi](x)=\frac{1}{(2\pi\h)^d}\int_{\R^d}e^{\frac{i}{\h}\xi\cdot(x-y)}h\left(\frac{x+y}{2},\xi\right)\varphi(y)\d y\d\xi
%\end{equation*}

%%%%%%%%%%%%%%%%%%%%%%%%%%%%%%%%%%%%%%%%%%%%%%%%%%%%%%%%
\section{Application to Evolution Equations \label{s:evol}}
%%%%%%%%%%%%%%%%%%%%%%%%%%%%%%%%%%%%%%%%%%%%%%%%%%%%%%%%

\begin{theo} \label{theo:main_result}
Let $e^{-\frac{i}{\eps} H^\eps t}$ be the propagator defined by the time-dependent Schr\"o\-dinger equation \eqref{eq:TDSE}
on the time-interval $[-T,T]$ with subquadratic potential $V\in C^\infty(\R^d,\R)$ and let $\Theta\in C^1([-T,T]);\Cc)$ be a time-dependent admissible spreading matrix such that $\Theta(t=0)$ is constant in $(y,\eta)$. Then
\begin{equation} \label{eq:main_estim}
\sup_{t\in [-T,T]}\left\|e^{-\frac{i}{\eps} H^\eps t}-\Ii^\eps\left(\kappa^t;\sum^N_{n=0}\eps^nu_n\right)\right\|_{L^2\to L^2}\leq C(T)\eps^{N+1},
\end{equation}
where $\kappa^t=(X^{\kappa^t},\Xi^{\kappa^t})$ and the $u_n(t,y,\eta)$ are uniquely given as
\begin{itemize}
\item the Hamiltonian flow associated with $h(x,\xi)=\frac{|\xi|^2}{2}+V(x)$ and
\item the solutions of the Cauchy-problems
\begin{eqnarray*}
\dt u_n(t,y,\eta) & = & \frac12{\rm tr}\left[\Y\left(F^\kt(y,\eta),\Theta\right)^{-1}\frac{d}{dt}\Y\left(F^\kt(y,\eta),\Theta\right)\right]u_n(t,y,\eta) \\
 & & +\sum_{k=0}^{n-1}L_{n,k}[\Theta]u_k(t,y,\eta)
\end{eqnarray*}
with initial conditions
\begin{eqnarray*}
u_0(0,y,\eta) & = & 1, \\
u_n(0,y,\eta) & = & 0, n\geq1.
\end{eqnarray*}
The matrix $\Y$ function of $F^\kt$, the Jacobian of $\kappa^t$, is defined in \eqref{eq:Y} and the $L_{n,k}$'s are differential operators whose coefficients are functions of derivatives of the Hamiltonian $h$ and of the admissible spreading matrix $\Theta$.
\end{itemize}
Moreover, for the Ehrenfest timescales $T(\eps)=C_T|\ln\eps|$, the constant $C(T)$ of \eqref{eq:main_estim} becomes $C'\eps^{-\rho}$ with $\rho$ arbitrary small if $C_T$ is taken small enough.
\end{theo}

\begin{rmqs} $ $
\begin{enumerate}
\item
The choice of the action associated to $\kappa^t$ is made by continuity in time with the initial condition $S^{\kappa^0}\equiv0$ and turns to be exactly the classical action integral defined by
\begin{equation*}
\int^t_0\left[\frac{d}{ds}X^{\kappa^s}(y,\eta)\cdot\Xi^{\kappa^s}(y,\eta)-h(\kappa^s(y,\eta))\right]\d s.
\end{equation*}
\item
The $L_{n,k}[\Theta]$'s can be obtained recursively by the process explained in the proof. Moreover, in the case where $\Theta$ does not depend on $(y,\eta)$, $L_{n,k}[\Theta]$ only depends on $n-k$ and involves only derivatives of $h$ of order between $n-k$ and $2(n-k)$, their formulae can be made rather explicit using the annihilation/creation point of view as presented in {\rm\cite{[SwartRousse]}} and {\rm\cite{[Swart]}}.
\end{enumerate}
\end{rmqs}

\begin{dem}
When $\Theta$ does not depend on $(y,\eta)$, the way to obtain recursive equations for the $u_n$'s was already explained in detail in \cite{[LaptevSigal]} and \cite{[Butler]} and strongly relies on Lemma~\ref{push}. Indeed, one easily gets, with $u^\eps=\sum^N_{n=0}\eps^n u_n$,
\begin{eqnarray*}
i\eps\dt\Ii^\eps(\kappa^t;u^\eps;\Theta(t)) & = & \Ii^\eps\left(\kappa^t;i\eps\dt u^\eps-\dt\Phi^\kt u^\eps;\Theta(t)\right) \\
H^\eps\Ii^\eps(\kappa^t;u^\eps;\Theta(t)) & = & \Ii^\eps\left(\kappa^t;\left[\frac{1}{2}|\nabla_x\Phi^\kt|^2-\frac{i\eps}{2}\Delta_x\Phi^\kt+V(x)\right]u^\eps;\Theta(t)\right). 
\end{eqnarray*}
As $\dt\Phi^\kt$, $|\nabla_x\Phi^\kt|^2$ and $\Delta_x\Phi^\kt$ are polynomial in $x$, one can remove this dependence using Lemma~\ref{push} and the identity $x=X^\kt+(x-X^\kt)$. To treat the $V(x)$ term, one uses Taylor formula at point $X^\kt$ to the $(2N+1)^{th}$ order and treat the polynomial appearing as before. As for the remainder, iterative applications of Lemma~\ref{push}, show that the gain in $\eps$ is at least $\eps^{N+2}$ with a symbol still dependent on $x$. Thus, we obtain
\begin{equation} \label{eq:remainder}
\left[i\eps\dt-H^\eps\right]\Ii^\eps(\kappa^t;u^\eps;\Theta(t))=\sum_{n=0}^{N+1}\eps^n\Ii^\eps(\kappa^t;v_n;\Theta(t))+\eps^{N+2}\Ii^\eps(\kappa^t;v^\eps_{N+2};\Theta(t))
\end{equation}
where the symbols on the right hand side are explicit functions of the $u_n$'s. More explicitly, we have
\begin{eqnarray*}
v_0 & = & \left[-\frac{d}{dt}S^{\kappa^t}+\frac{d}{dt}X^{\kappa^t}\cdot\Xi^{\kappa^t}-h\circ\kappa^t\right]u_0 \\
v_{n+1} & = & i\frac{d}{dt}u_n-\frac{i}{2}{\rm Tr}\left(\Y^{-1}\frac{d}{dt}\Y\right)u_n-\sum_{k=0}^{n-1}iL_{n,k}[\Theta]u_k \qquad (0\leq n\leq N)
\end{eqnarray*}
where the $L_{n,k}[\Theta]$'s are linear differential operators with coefficients in $S[0,2d]$ (more precisely, they are polynomials in derivatives of the flow $\kt$ and in derivatives, of at least second order, of the potential $V(x)$). The cancellation of the prefactor in $v_0$ (we do not want $u_0$ to vanish identically) shows that $\kt$ must be the flow of $h$. Next, cancellations of the $v_n$'s for $1\leq n\leq N+1$ give the transport equations for the $u_n$'s whose solutions are easily seen to be bounded. The final step is then to check that the remaining symbol $v^\eps_{N+2}$ is bounded so that Theorem~\ref{theo:L2bound} applies.

For non-constant admissible spreading matrix, the proof follows the same lines. However, we need to reformulate Lemma~\ref{push} because we have now
\begin{equation*}
\left(v\cdot\nabla_\eta\right)\Phi^\kappa=v\cdot\Y(F^\kappa;\Theta)(x-X^\kappa)+\frac{i}{2}(x-X^\kappa)\cdot\left[\left(v\cdot\nabla_\eta\right)\Theta\right](x-X^\kappa)
\end{equation*}
so that the lemma involved is now

\begin{lem}
Let $u\in S[+\infty,3d]$ and $V(x,y,\eta)$ a $\C^d$-valued function in $S[+\infty,3d]$, then
\begin{eqnarray}
\lefteqn{\Ii^\h\left(\kappa;V(x,y,\eta)\cdot(x-X^\kappa(y,\eta))u;\Theta\right)=} \label{eq:IPPspread} \\
 & & i\h \Ii^\h\left(\kappa;L(\kappa;\Theta;V)u;\Theta\right)+\frac{i}{2}\Ii^\h\left(\kappa;(x-X^\kappa)\cdot{\Y(F^\kappa;\Theta)^\dagger}^{-1}\left[\left(V\cdot\nabla_\eta\right)\Theta\right](x-X^\kappa)u;\Theta\right) \nonumber
\end{eqnarray}
where $L(\kappa;\Theta;V)$ is defined in \eqref{IPP}.
\end{lem}

Thus, one removes one $(x-X^\kappa)$ (gaining one order in $\eps$) to the price of adding a quadratic term in $(x-X^\kappa)$. To summarize, either we gain in orders of $\eps$ or either we produce polynomials in $(x-X^\kappa)$ of higher degree. Thus, a straightforward induction shows that there exist linear differential operators $L_{n,\alpha}[\Theta]$ and $L_{k,\alpha,\beta}[\Theta]$ such that
\begin{eqnarray*}
\Ii^\h\left(\kappa;V_\alpha(x,y,\eta)(x-X^\kappa(y,\eta))^\alpha u;\Theta\right) & = & \sum^{N+1}_{n=\lceil\frac{|\alpha|+1}{2}\rceil}\hspace{-0.3cm}\h^n\Ii^\h\left(\kappa;L_{n,\alpha}[\Theta](V_\alpha u);\Theta\right) \\
 & & +\sum_{k+\frac{|\beta|}{2}\geq N+2}\hspace{-0.3cm}\h^k\Ii^\h\left(\kappa;(x-X^\kappa)^\beta L_{k,\alpha,\beta}[\Theta](V_\alpha u);\Theta\right)
\end{eqnarray*}

To estimate the remainder, one combines the polynomial $(x-X^\kappa)^\beta$ with the Gaussian part of $\exp\left(\frac{i}{\eps}\Phi^\kappa\right)$ to obtain the appropriate gain in orders of $\eps$ (see Section~\ref{s:spread} above and Section~7 of \cite{[Swart]}):

\begin{lem}
Let $u\in S[+\infty,3d]$, then
\begin{equation*}
\left\Vert\Ii^\h\left(\kappa;(x-X^\kappa)^\beta u;\Theta\right)\right\Vert\leq C_{\beta}[\Theta_0]\h^{|\beta|/2}\sum_{\substack{\alpha_j=0,1 \\ \beta_j=0,1 \\ |\gamma|\leq d}}\Vert\partial^\alpha_x\partial^\beta_y\partial^\gamma_\eta u\Vert_\infty.
\end{equation*}
\end{lem}

Finally, for the Ehrenfest timescales, it is enough to see, using estimates from \cite{[Bambusi]} or \cite{[Bouzouina]}, that the explicit forms of the $u_n$'s for $n\geq2$ and thus of the symbol in the remainder~\eqref{eq:remainder} involve derivatives of the potential (of at least second order) which are assumed to be globally bounded and derivatives of the linearized flow $F^\kt$ which have uniform bounds $C'\eps^{-\rho}$ as already stated in Propostion~$1$ of \cite{[SwartRousse]}.
\end{dem}

\begin{rmq} \label{rk:toPDO}
The fact that our class of FIOs is not stable under the action on the left of pseudodifferential operators (some of them, though bounded, can not be represented as FIOs with $\kappa={\rm id}$ and a bounded symbol) makes the generalization of this Theorem to general Hamiltonians rather technical. However, following the lines of Section~5 in {\rm\cite{[Swart]}}, it should not be out of reach.
\end{rmq}

%%%%%%%%%%%%%%%%%%%%%%%%%%%%%%%%%%%%%%%%%%%%%%%%%%%%%%%%
\appendix
%%%%%%%%%%%%%%%%%%%%%%%%%%%%%%%%%%%%%%%%%%%%%%%%%%%%%%%%

%%%%%%%%%%%%%%%%%%%%%%%%%%%%%%%%%%%%%%%%%%%%%%%%%%%%%%%%
\section{Oscillatory Integral with Complex Phase \label{oscill}}
%%%%%%%%%%%%%%%%%%%%%%%%%%%%%%%%%%%%%%%%%%%%%%%%%%%%%%%%

We present here the standard machinery of oscillatory integrals. For the definition of expressions like
\begin{equation} \label{osc}
\frac{1}{(2\pi\h)^d}\int_{\R^{2d}}e^{\frac{i}{\h}\Phi(x,y,\eta)}a(x,y,\eta)d\eta\d y,
\end{equation}
which have no sense as an ordinary Lebesgue-integral, two approaches can be taken. First, one can choose a function $\sigma\in\Ss(\R^d)$ with $\sigma(0)=1$ and set
\begin{equation*}
\eqref{osc}:=\lim_{\lambda\to+\infty}\frac{1}{(2\pi\h)^d}\int_{\R^{2d}}\sigma(\eta/\lambda)e^{\frac{i}{\h}\Phi(x,y,\eta)}a(x,y,\eta)d\eta\d y.
\end{equation*}
To show the independence of the function $\sigma$ a second technique is required. Under suitable conditions on the phase function, see for instance \cite{[Martinez]}, the operator
\begin{equation*}
L=\frac{1}{1+|\Bb(y,\eta)\nabla_{(y,\eta)}\Phi(x,y,\eta)|^2}\left[1-i\h\overline{\Bb(y,\eta)\nabla_{(y,\eta)}\Phi(x,y,\eta)}\cdot\Bb(y,\eta)\nabla_{(y,\eta)}\right]
\end{equation*}
provides decay in $\eta$ by partial integrations, {\it i.e.}
\begin{equation*}
\left|(L^\dagger)^ku\right|\leq\frac{M_k}{(1+|\eta|^2)^{k/2}}\sum_{|\alpha|\leq k}|\partial^\alpha_y u|,
\end{equation*}
where $L^\dagger$ is the symmetric of $L$ defined by
\begin{equation*}
\int(L\varphi)(y)\psi(y)\d y=\int\varphi(y)(L^\dagger\psi)(y)\d y, \qquad \forall\varphi,\psi\in\Ss(\R^d).
\end{equation*}
Hence an alternative definition is provided by
\begin{equation*}
\eqref{osc}=\frac{1}{(2\pi\h)^d}\int_{\R^{2d}}e^{\frac{i}{\h}\Phi(x,y,\eta)}\left(L^\dagger\right)^ka(x,y,\eta)d\eta\d y.
\end{equation*}
For the special case of the phase function $\Phi^\kappa$, the operator $L$ reads
\begin{equation*}
L=\frac{1+i\h\eta\cdot\left[\nabla_y+\Aa(y,\eta;\Theta)\nabla_\eta\right]}{1+|\eta|^2},
\end{equation*}
where $\Aa(y,\eta;\Theta)$ is defined in \eqref{Aa}, and provides the expected decay in the $\eta$-variable. As for the decay in the $y$ variable, it comes from the Schwartz class of $\varphi$ in $a(x,y,\eta)=u(x,y,\eta)\varphi(y)$.

%%%%%%%%%%%%%%%%%%%%%%%%%%%%%%%%%%%%%%%%%%%%%%%%%%%%%%%%
\section{The Short-Time Fourier Transform \label{Wigner}}
%%%%%%%%%%%%%%%%%%%%%%%%%%%%%%%%%%%%%%%%%%%%%%%%%%%%%%%%

We begin with a definition.
\begin{defi} \label{def:stft}
For $f$ and $g$ two functions in the Schwartz space $\Ss(\R^d)$, we define their {\bf short-time Fourier transform} as the function on $\R^d\times\R^d$
\begin{equation*}
V_g[f](y,\eta)=\frac{1}{(2\pi)^{d/2}}\int_{\R^d}e^{-i\eta\cdot x}f(x)\overline{g(x-y)}\d x.
\end{equation*}
\end{defi}

\begin{lem}
The bilinear operator $(f,g)\in\Ss(\R^d)\times\Ss(\R^d)\mapsto V_g[f]$ extends by continuity to an operator $L^2(\R^d)\times L^2(\R^d)\to L^2(\R^d\times\R^d)$ and
\begin{equation*}
\Vert V_g[f]\Vert_{L^2}=\Vert g\Vert_{L^2}.\Vert f\Vert_{L^2}.
\end{equation*}
\end{lem}

\begin{dem}
We use Parseval formula in $\eta$ and Fubini theorem to get
\begin{equation*}
\Vert V_g[f]\Vert_{L^2}^2
=\Vert\Ff^{-1}_\eta V_g[f]\Vert_{L^2}^2
=\Vert f(x)\overline{g(x-y)}\Vert_{L^2}^2
=\int_{\R^d}|f(x)|^2\left(\int_{\R^d}|g(x-y)|^2\d y\right)\d x.
\end{equation*}
Hence, the result.
\end{dem}

$ $

The reader interested in more properties of this transform should consult \cite{[Groechenig]}. Anyway, we extend here the definition of short-time Fourier transform for window functions $g$ Gaussian in $x$ but also dependent on the variables $y$ and $\eta$, more precisely of the form
\begin{equation} \label{wind}
g(x,y,\eta)=G(x)\tilde{g}(x,y,\eta)
\end{equation}
with $G(x)=\exp(-x\cdot\Theta x/2)$ and $\tilde{g}\in S[(+\infty,0);(d,2d)]$%(in fact it is enough to assume polynomial bound for the derivatives $\partial^\alpha_x\tilde{g}$ with $\alpha_j=0,1$)
, by the formula
\begin{equation} \label{stft2}
\tilde{V}_g[f](y,\eta)=\frac{1}{(2\pi)^{d/2}}\int_{\R^d}e^{-i\eta\cdot x}f(x)\overline{g(x-y,y,\eta)}\d x.
\end{equation}
The preceeding lemma also extends to this situation and states the following.

\begin{lem}
If $g$ is of the form \eqref{wind}, then the linear operator $f\mapsto \tilde{V}_g[f]$ extends by continuity to $L^2(\R^d)\to L^2(\R^d\times\R^d)$ and there exists a constant $C[\Theta]>0$ such that, for any $f\in L^2(\R^d)$,
\begin{equation*}
\left\Vert\tilde{V}_g[f]\right\Vert_{L^2}\leq C[\Theta]\left(\sum_{\alpha_j=0,1}\Vert\partial^\alpha_x\tilde{g}\Vert_{L^\infty}\right)\Vert f\Vert_{L^2}
\end{equation*}
where the sum runs on all multi-indices $\alpha=(\alpha_1,\ldots,\alpha_d)$ such that any $\alpha_j=0$ or $1$.
\end{lem}

\begin{dem}
We write $\left\Vert\tilde{V}_g[f]\right\Vert_{L^2}^2$ as the integral in $(x^1,x^2,\xi^1,\xi^2,y,\eta)$ of
\begin{equation*}
e^{i(\xi^1-\eta)\cdot x^1}(\Ff f)(\xi^1)\overline{g(x^1-y,y,\eta)}e^{-i(\xi^2-\eta)\cdot x^2}\overline{(\Ff f)(\xi^2)}g(x^2-y,y,\eta).
\end{equation*}
Integrating by parts with respect to $x^1$ and $x^2$ turns the integrand into sum of terms
\begin{equation*}
\frac{e^{i(\xi^1-\eta)\cdot x^1}}{\displaystyle\prod_{j=1}^d(1-i(\xi^1_j-\eta_j))}(\Ff f)(\xi^1)\overline{g_1(x^1-y,y,\eta)}\frac{e^{-i(\xi^2-\eta)\cdot x^2}}{\displaystyle\prod_{j=1}^d(1+i(\xi^2_j-\eta_j))}\overline{(\Ff f)(\xi^2)}g_2(x^2-y,y,\eta)
\end{equation*}
with $g_1$ and $g_2$ of the form \eqref{wind} and whose integral reads
\begin{equation*}
\int_{\R^{4d}}e^{i\eta\cdot(x^2-x^1)}V_{p_0}[\Ff f](-x^1,\eta)\overline{V_{p_0}[\Ff f](-x^2,\eta)}g_2(x^2-y,y,\eta)\overline{g_1(x^1-y,y,\eta)}\d x^1\d x^2\d y\d\eta
\end{equation*}
where $p_0$ stands for the $L^2$ function defined in \eqref{p}.
Moreover, the Gaussian parts of $\overline{g_1}$ and $g_2$ combine into
\begin{equation*}
\hspace{-.1cm}\exp\left[-\left|(\Re\Theta)^{1/2}\left(\frac{x^1+x^2}{2}-y\right)\right|^2-\frac{\left|(\Re\Theta)^{1/2}(x^1-x^2)\right|^2}{4}+i(x^1-x^2)\cdot\Im\Theta\left(\frac{x^1+x^2}{2}-y\right)\right]
\end{equation*}
so that $\left\Vert\tilde{V}_g[f]\right\Vert_{L^2}^2$ is a finite sum of integral in $(x^1,x^2,y,\eta)$ of terms of the form
\begin{equation*}
\psi_1(x^1,\eta)\psi_2(x^2,\eta)y^\alpha e^{-|(\Re\Theta)^{1/2}y|^2}(x^1-x^2)^\beta e^{-\frac{1}{2}|(\Re\Theta)^{1/2}(x^1-x^2)|^2}\psi_3(x^1-y,x^2-y,y,\eta)
\end{equation*}
with $\psi_1,\psi_2\in L^2(\R^{2d};\C)$, $\psi_3\in S[(+\infty,0),(2d,2d)]$ and $\alpha,\beta\in\N^d$.
As estimating those integral, up to raising $\alpha$ and $\beta$ to take into account the polynomial growth of $\psi_3$, we end up with, after the change of variables $(x^1,x^2)=(x,x-z)$,
\begin{eqnarray*}
\lefteqn{\hspace{-.8cm}\Vert\psi_3\Vert_{L^\infty}\left(\int_{\R^d}|y^\alpha|e^{-|(\Re\Theta)^{1/2}y|^2}\d y\right)\int_{\R^d}\left(\int_{\R^{2d}}|\psi_1(x,\eta)\psi_2(x-z,\eta)|\d x\d\eta\right)|z^\beta|e^{-\frac{1}{2}|(\Re\Theta)^{1/2}z|^2}\d z} \\
 & \leq & \Vert\psi_3\Vert_{L^\infty}\left(\int_{\R^d}|y^\alpha|e^{-|(\Re\Theta)^{1/2}y|^2}\d y\right)\left(\int_{\R^d}|z^\beta|e^{-\frac{1}{2}|(\Re\Theta)^{1/2}z|^2}\d z\right)\Vert\psi_1\Vert_{L^2}\Vert\psi_2\Vert_{L^2}
\end{eqnarray*}
We conclude using $\Vert\psi_1\Vert_{L^2}=\Vert\psi_2\Vert_{L^2}=\pi^{d/2}\Vert f\Vert_{L^2}$ (because of $\Vert p_0\Vert_{L^2}=\pi^{d/2}$).
\end{dem}

\noindent{\bf Acknowledgements} : The author would like to thank C. Fermanian Kammerer for fruitful discussions and J. Le Rousseau for indicating him the strategy of Hwang.

\noindent Vidian \textsc{Rousse}, Universit\'e Paris Est, 
UFR des Sciences et Technologie,
61, avenue du G\'en\'eral de Gaulle,
94010 Cr\'eteil Cedex, France.\\
{\tt vidian.rousse@univ-paris12.fr}

\end{document}